\documentclass[12pt]{book}
\usepackage{wrapfig,epsfig,color}
\usepackage{makeidx,tsukuba}

\makeauthorindex
\makeindex

\begin{document}
\BookTitle{\itshape The 28th International Cosmic Ray Conference}
\CopyRight{\copyright 2003 by Universal Academy Press, Inc.}
\pagenumbering{arabic}

\chapter{Computational Techniques for Simulating Light Propagation in High-Energy Neutrino Telescopes}

\author{Predrag Mio\v{c}inovi\'c$^1$ and Peter Nie{\ss}en$^2$ \\
{\it(1) Department of Physics, University of California, Berkeley, CA 94720, USA \\
(2) Vr\"{y}e Universiteit Brussel, Dienst ELEM, B-1050 Brussel, Belgium }
}

\section*{Abstract}
To maximize the accuracy of background simulation and event reconstruction, 
high-energy neutrino telescopes require detailed knowledge of light propagation
over a large volume of detection medium. If light scattering and absorption lengths
in the medium are of the same scale as the detector size, this problem can only be 
handled numerically. Any inhomogeneity of optical properties in the
medium further complicates the problem, requiring large computational resources.
We present a treatment based on combining ray-tracing Monte Carlo and neural 
network techniques which offers a reasonable compromise between solution 
accuracy, computer memory and CPU usage.

\section{Introduction}

Any high-energy neutrino telescope has to reject an overwhelming background 
produced by cosmic-ray-generated atmospheric muons before any neutrino signal
can be extracted. The background data
simulation needs to faithfully reproduce all known classes of background 
events, while the event reconstruction should be accurate enough to allow their 
separation from the expected signal. For these tasks, correctly describing 
the photon transport in the detector is crucial.

An important requirement for a functional neutrino telescope is that its size 
should be smaller than the volume at which photon transport through its 
detection medium can be considered diffusive, but also large enough to 
have reasonable signal collection area. In a medium where absorption dominates
over scattering, like deep ocean or lake sites, the light transport problem can be
solved analytically if scattering is neglected or treated as a correction. 
In the regime where scattering and absorption
distances are similar, as is the case for AMANDA~[4], the problem has
to be tackled numerically. To satisfy the above requirement, the numerical
solutions have to be tabulated over large volumes.

In this report we describe numerical techniques that were used to produce 
satisfactory results for use in AMANDA in terms of solution accuracy and
ease of implementation into computational structure. The same approach can 
be easily adapted to any other detector facing similar problems. 

\section{Photon table generation}
The distribution of time dependent photon fluxes around a light source is
generated by a photon transport Monte Carlo simulation which has been
optimized for numerical accuracy and speed of execution~[3]. 
The photon transport is done in a fashion analogous to the ray-tracing technique 
employed in computer graphics design. The main difference is that instead of
recording the ``illumination'' of the predefined set of objects, the entire
photon propagation volume is subdivided into cells which independently 
record time-dependent photon flux. This method allows a rapid collection 
of large photon statistics: $\sim\!10^6\;\gamma$/CPU~hour on 650~MHz Pentium~III.

Since the detector medium is considered not to have boundaries, 
light scattering is caused only by impurities contained within the ice, which
also dominate the absorption in the visible wavelength range. The depth-dependent 
concentration and composition of these impurities has been 
studied and a model for the description of optical properties
of ice in AMANDA as a function of depth has been developed~[5]. 
The inhomogeneity of optical properties of ice partially breaks the 
translational and rotational symmetries of the problem and requires that each 
light emission point be treated separately. 

In AMANDA, the instrumentation is surrounded by columns of re-frozen ice enriched 
in microscopic air-bubbles. This ice occupies only a very small 
fraction of the detection volume, and will only affect
photon distribution near the detection point. Thus it
can be treated as a perturbation to the directional sensitivity of the
light detectors instead of being incorporated into the photon 
transport simulation.
\begin{figure}[!b]
  \hfill
  \begin{minipage}[t]{.49\textwidth}
    \begin{center}  
      \epsfig{file=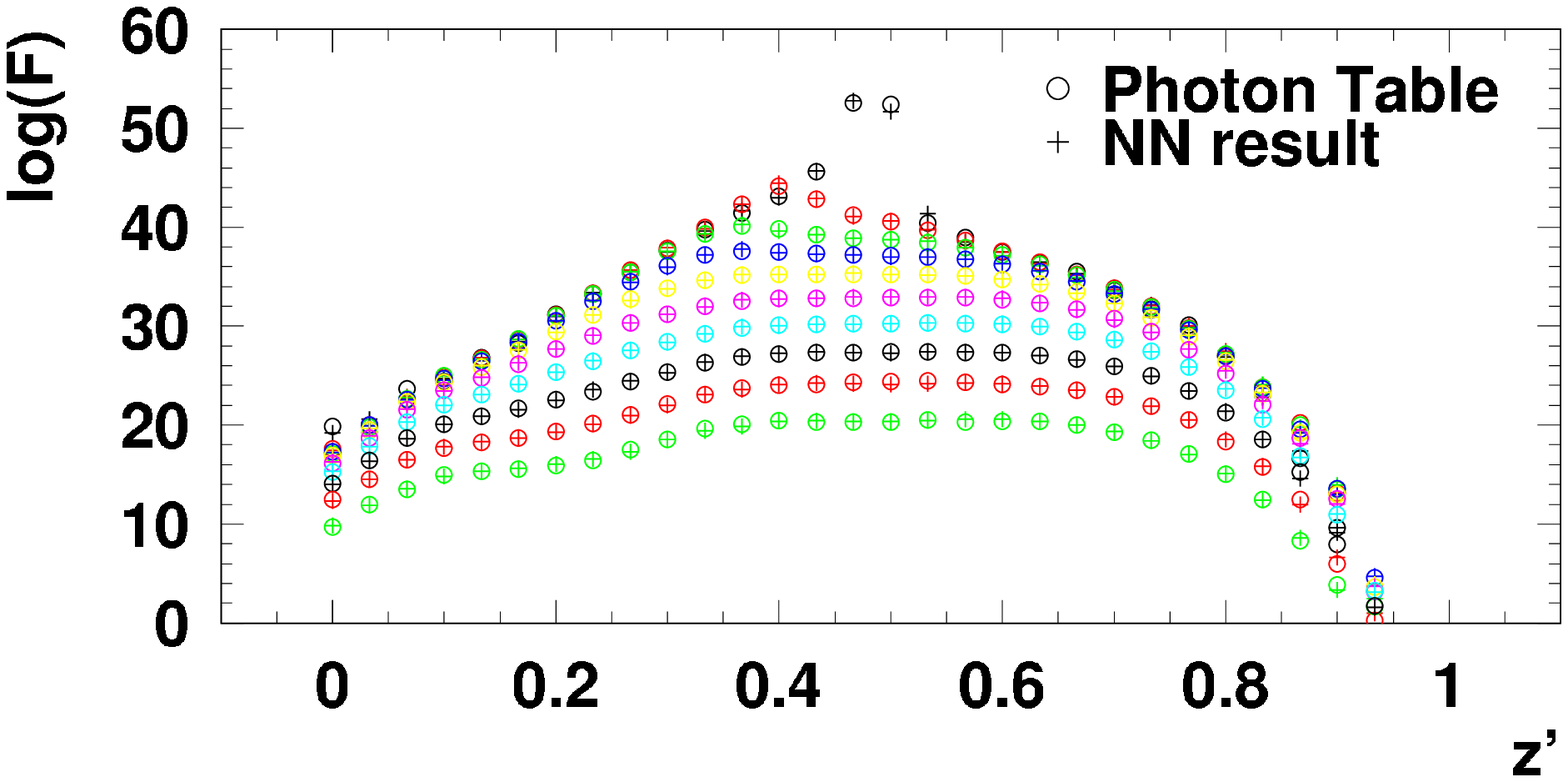, scale=0.35}
      \caption{Logarithm of the photon fluence
$\mathcal{F}$ for points along the particle track $z^\prime$ at distances
$\rho$ = 0, 3, 12, 27, 48, 75, 108, 147,
192 and 243 meters from the track -- top to bottom.}
      \label{AmpFit}
    \end{center}
  \end{minipage}
  \hfill
  \begin{minipage}[t]{.49\textwidth}
    \begin{center}  
      \epsfig{file=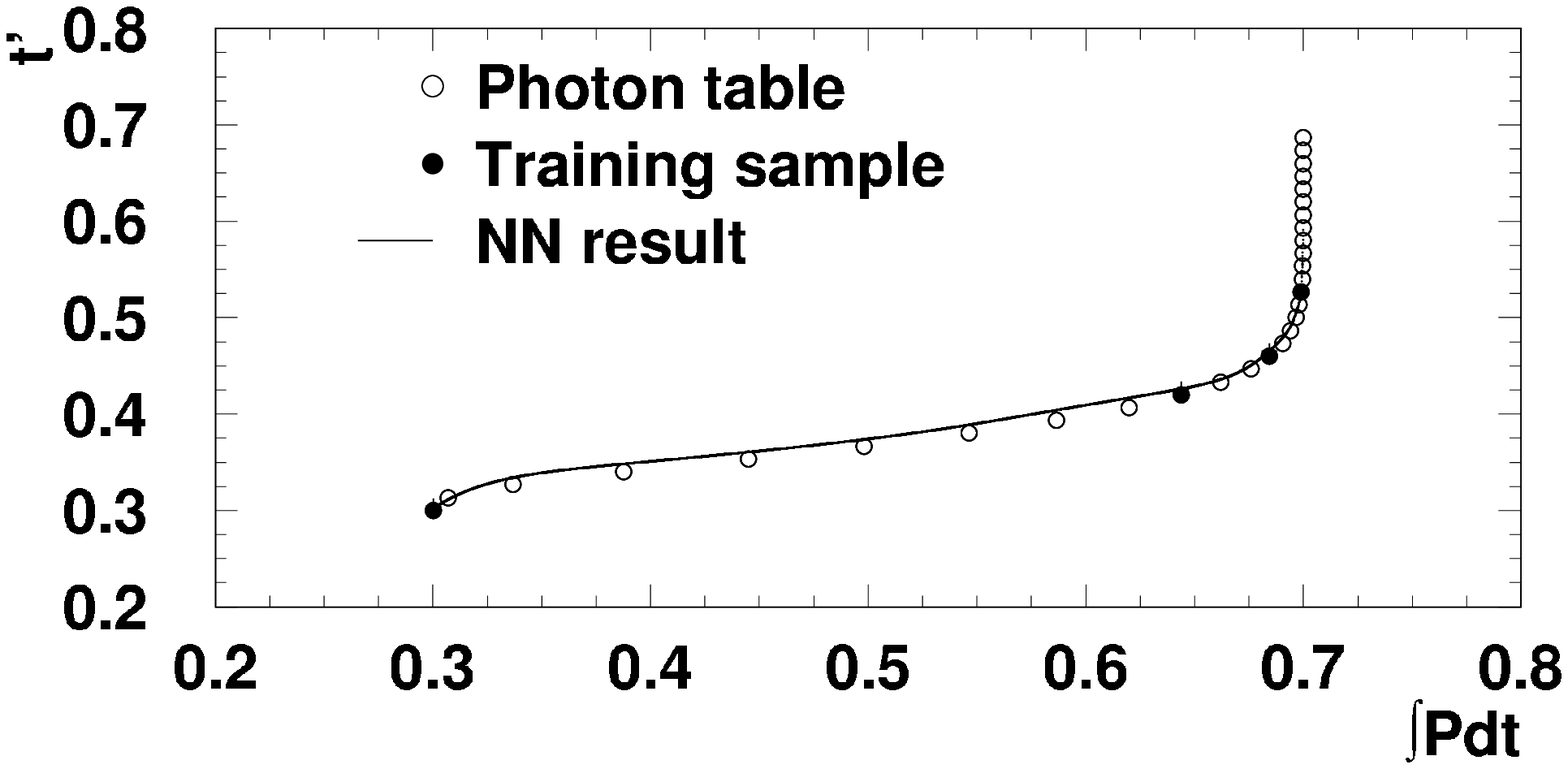, scale=0.35}
      \caption{ An example of how the network interpolates the integrated and  
	inverted time-delay probability tables. The axes units are rescaled 
	for use in~NN.}
      \label{TimeExample}
    \end{center}
  \end{minipage}
  \hfill
\end{figure}

For the purpose of detector response simulation, photon distributions resulting from
all relevant source depths and orientations have to be simulated. In the case
of AMANDA, this covers $\sim$700 meters of depth and 4$\pi$ of solid angle. 
Combining the sizes of tables describing the photon flux around each of 
these sources, the overall size of the photon table set becomes very large;
$\sim$1 GB even for a very coarse binning.
Increasing the quality (for improved simulation accuracy)
or the number (for larger detectors~[1]) of 
tables quickly becomes prohibitive if one should hope to use most readily-available 
computational resources. 

Possible solutions would be to either segment detector simulation in such a way that
only a subset of all tables would be needed at any given time, or to reduce 
the memory needed to describe the content of each table. 
To pursue the second option, we 
have chosen to use neural networks to make a 
model-free fit to the tables and use the network output in the 
detector simulation. 

\section{Neural Network (NN) implementation}

The photon flux is stored in tables as photon fluence
$\mathcal{F}_{ij}=f(\mathbf{r}_{\mathrm{src},i}, \mathbf{r}_j)$ at a point $\mathbf{r}_j$
due to a source at $\mathbf{r}_{\mathrm{src},i}$, and as a normalized time 
profile of the flux
$P_{ijk}=g(\mathbf{r}_{\mathrm{src},i}, \mathbf{r}_j, t_k)$ where
$t_k$ is time delay with respect to an unscattered photon 
from the same source. We now represent the functions $\mathcal{F}$ and $P$ by 
multi-layer perceptron (MLP) neural networks~[2] with as many input nodes as 
there are coordinates, a single output node for the function value, and 
a number of hidden nodes to be determined. The table coordinates and
stored values are used to create neural net training patterns.  

The photon tables map  5 spatial coordinates
to a fluence which can extend over 20 orders of magnitude in the case of
AMANDA. This
is beyond the dynamic range of a neural net, so the logarithm of the
fluence is fitted. All input coordinates are rescaled on [0,1] range. 
In order to find a suitable net architecture, we consider
first a fluence projection onto the dimension along which the function
shows the most features.
Hidden nodes and layers are added/removed until we find a satisfactory 
agreement between the neural net output and the tabulated value
expressed in the {\em linear} fluence units. After this, input nodes are 
added one at a time for the additional, smoother dimensions. The hidden layer
configuration can be modified if necessary, but we usually find that the
additional
links from the new input nodes can handle the additional dimensions. We
train the net
until the difference between the network output and the desired
function values stabilizes. Figure \ref{AmpFit} shows the excellent network
response to the fluence training for the light emitted by a short
particle-track segment. 

Fitting delay tables follows the same method as the fluence
tables, but with one additional input coordinate.
In the detector simulation, one wishes to randomly sample 
$P$ in order to generate timing response of the detector. To do this, one is 
interested in the integrated and time-inverted function of $P$,
\mbox{$t=h(\int_0^t\!P dt^\prime)$}, 
which can be easily expressed in neural net formulation. 
If the function $\int_0^t\!P dt^\prime$ is not {\it one-to-one}, we
exclude the flat part of the function from the fit, and if due to binning effects
$t\ne 0$ for all values of $\int_0^t\!P dt^\prime$, we add a 
pattern \mbox{$(\int_0^t\!P dt^\prime=0, t=0)$} to make the function {\it on-to}.

\begin{wrapfigure}[20]{r}{0.5\columnwidth}
\begin{center}
\epsfig{file=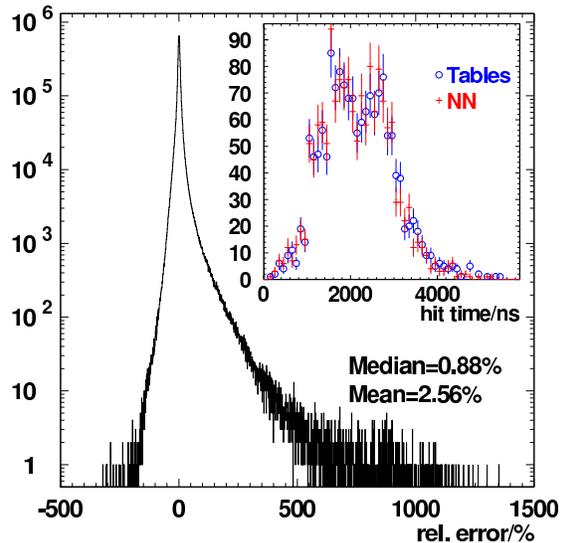,width=0.50\columnwidth}
\caption{The relative time-delay error of NN fitted photon flux. 
{\bf Inset:} Hit time distribution as simulated for muons passing
AMANDA, comparing table lookup and NN approach.}
\label{TimeFit}
\end{center}
\end{wrapfigure}

To avoid biasing the network to the order in which patterns are
stored, they are shuffled before network training. For network
construction and training we use the SNNS package~[6]. 
In the case of fluence tables,
all patterns fit into a standard computer memory and can be used in a single training
session.
In the case of delay tables, only $\sim$10\% of patterns fit into memory, 
so a random subsample is
used. The network's interpolation capability (Fig. \ref{TimeExample})
ensures that the patterns not used for training are nevertheless accurately
reproduced within 2.6\% (Fig. \ref{TimeFit}).
Large relative-error tails, seen in Fig. \ref{TimeFit}, occur only for very short 
time-delays and produce no adverse effect on application of net output. 

\section{Results}

After training, the network can be used in the standard detector response
simulator. To check the accuracy of the net, we compare events
simulated using the direct table lookup to events simulated using the
neural net representation. It is found that simulated hit-amplitude and hit-time 
distributions are in good agreement (inset Fig. \ref{TimeFit}).
The memory reduction factor achieved is $\sim$1000 for the fluence tables and
$\sim$0.5$\times 10^6$ for the timing tables. We observe no CPU runtime 
increase due to NN evaluation.

We would like to thank K. Woschnagg, S. Hundertmark, L. Gerhardt, and 
M. Kowalski for help and advice.

\section*{References}
\re
1. Ahrens~J. 2003, ArXiv: astro-ph/0305196
\re
2. Haykin~S. 1999, in ``Neural Networks: A Comprehensive Foundation'' (Prentice-Hall, New Jersey)
\re
3. Mio\v{c}inovi\'c~P. 2001, Ph.D. thesis, {\tt http://area51.berkeley.edu/manuscripts}
\re
4. Wagner~W. et al., these proceedings
\re 
5. Woschnagg~K. 1999, in Proc. 26$^\mathrm{th}$ ICRC (IUPAP, Salt Lake City), Vol. 2, 200
\re
6. Zell~A. et al. 1995, SNNS User Manual, IPVR, Univ. of Stuttgart, Germany

\endofpaper
\end{document}